\documentclass[a4paper]{article}
\usepackage{Odyssey2022}
\usepackage{epsfig,amssymb,amsmath}
\usepackage{cite}
\usepackage{float}
\usepackage{hyperref}
\usepackage{balance,subcaption,xcolor}
\usepackage{tipa}
\hypersetup{
    colorlinks=true,
    linkcolor=purple,
    filecolor=magenta,      
    urlcolor=black,
}
\ninept

\title{Explainable deepfake and spoofing detection: \\ an attack analysis using SHapley Additive exPlanations}

\makeatletter
\def\name#1{\gdef\@name{#1\\}}
\makeatother
\name{{\em Wanying Ge, Massimiliano Todisco and Nicholas Evans}}

\address{EURECOM, Sophia Antipolis, France \\
{\small \tt lastname@eurecom.fr} }
\begin{document}
\maketitle

\begin{abstract}

Despite several years of research in deepfake and spoofing detection for automatic speaker verification, little is known about the artefacts that classifiers use to distinguish between bona fide and spoofed utterances. An understanding of these is crucial to the design of trustworthy, explainable solutions. In this paper we report an extension of our previous work to better understand classifier behaviour to the use of SHapley Additive exPlanations (SHAP) to attack analysis. Our goal is to identify the artefacts that characterise utterances generated by different attacks algorithms. Using a pair of classifiers which operate either upon raw waveforms or magnitude spectrograms, we show that visualisations of SHAP results can be used to identify attack-specific artefacts and the differences and consistencies between synthetic speech and converted voice spoofing attacks.

\end{abstract}

\vspace{0.1cm}
\section{Introduction}
    
    Deep neural networks (DNNs) are now the de facto approach to spoofing detection for automatic speaker verification~\cite{lavrentyeva2019stc_lcnn,peddinti2015_tdnn, Lai2019_assert, Park2019freqmask}. Spoofing countermeasures usually operate upon hand-crafted spectro-temporal features extracted from the input utterance~\cite{nautsch2021asvspoof, li2020res2net,zhang2021oneclass, chen2020generalization, tomilov21_asvspoof, chen21_URchannel}. Lately, so-called raw end-to-end (E2E) solutions~\cite{tak2021rawnet2, ge21_raw_pc_darts, hua2021TSSDNet, tak21_gat, Jung2021_AASIST} are beginning to outperform the more traditional approaches. By operating directly upon raw audio waveforms, such systems have greater potential to capture the tell-tale signs of spoofing attacks, e.g.\ speech synthesis, converted voice and replay, the artefacts of which might not be captured in handcrafted features that are often designed for other tasks. 
    
    Whatever the approach, there is usually an interest to understand \emph{how} or \emph{why} a given classifier arrives at the scores or the decisions it produces.  Such an understanding is crucial to the design of trustworthy systems and key to the trend toward explainable artificial intelligence (xAI)~\cite{arrieta2020_xai,  adadi2018_peeking}.  So far, while we know that different spoofing detection solutions can rely upon different temporal or spectral intervals~\cite{chettri2018_analysing,tak2021rawnet2, zhang21da_cas}, we know surprisingly little else, despite several years of research.
    
    An explanation of the cues used by a classifier in distinguishing between bona fide and spoofed speech is of interest not just in terms of curiosity.  Explanations can be crucial to some particular use case scenarios, e.g.\ those involving forensics, but they might also reveal opportunities to improve performance through feature engineering and/or classifier design.  Explanations might also be particularly relevant in the case of now-dominant deep neural networks, increasingly opaque, black-box spoofing detection solutions with no obvious means to help explain classifier behaviour, scores or decisions.
    
    Motivated by related explainability studies in other speech tasks~\cite{becker2018interpreting,muckenhirn19_visualiseCNN,sivasankaran21_SHAP_SE},
    we have investigated the use of SHapley Additive exPlanations (SHAP)~\cite{SHAP} to explore and compare the behaviour of DNN-based solutions to spoofing detection~\cite{ge2021SHAP}.  SHAP is a feature attribution approach to explainability.  A so-called SHAP value is calculated and assigned individually to each and every input feature.  For image-related tasks, the feature can be an image pixel.  For speech-related tasks, the feature can be a spectro-temporal magnitude estimate or, as we show later in this paper, a sample of the input time-domain waveform. 
    SHAP values reflect the difference in classifier output when derived either with or without the use of each feature.  The SHAP value hence indicates the influence or importance of each input feature upon the classifier output. 
    
    SHAP values can be readily visualised with heat maps of the same dimension as the input image or spectro-temporal decomposition.  We followed this approach in our previous work to explore differences in \emph{classifier behaviour} and confirmed that different solutions exploit different temporal intervals and spectral sub-bands to distinguish between bona fide and spoofed utterances, behaviour that cannot be derived from classifier scores or decisions alone.  The current article reports our work to extend the study to \emph{attack analysis}.  
    
    Here we are interested to learn more about the attacks themselves, namely the specific artefacts that characterise each attack or class of attacks, e.g., algorithms such as text-to-speech (TTS) or voice conversion (VC). We focus upon the training data in the ASVspoof 2019 Logical Access (LA) database and use a pair of classifiers trained using spoofed data generated with specific attack algorithms to identify the characteristic artefacts. We found that this approach is critical to the identification of consistent artefacts in training data for which ground truth attack algorithms are available. This same approach, though, is obviously impracticable in the case of evaluation data, or spoofing detection in the wild, where the ground truth label of the spoofing attack algorithm is, by default, unknown. In this case, explainability proves to be considerably more challenging.
    Nonetheless, there are similarities between the characteristics of different known attacks, similarities which provide a starting point for explainability in the wild. 
    
    The new contributions in this work are: (i)~the first application of SHAP analysis to an end-to-end spoofing detection solution which operates upon raw waveform inputs; (ii)~an attack analysis in terms of explainability using two different classifiers, one that operates upon spectro-temporal magnitude estimates, the other upon raw waveform inputs; (iii)~explainability visualisations for both classifier inputs.

\vspace{0.3cm}
\section{SHapley Additive exPlanations (SHAP)}

    We provide a brief description of SHAP analysis.  In extending our previous work~\cite{ge2021SHAP}, we also show how it can be applied to raw, temporal inputs and provide example visualisations applied to classifiers that operate upon either spectro-temporal decompositions or raw waveforms.

    \subsection{Definition}

        SHAP~\cite{SHAP} is used to explain the contribution of each individual input feature to the output of a given model $f(x)$, here a spoofing detection classifier.  The contribution of each feature is reflected via SHAP values $\phi$, a reflection of the difference in the model output when it is learned with or without the inclusion of this particular feature. When $f(\mathbf{x})$ is a complex model which takes considerable time to retrain, e.g.\ deep neural network solutions, some simplifications are necessary to mitigate the need for repetitive retraining. First, the input $\mathbf{x}$ is approximated by a simplified feature $\mathbf{x}'=\{x'_{1}, ..., x'_{D}\}$, where $x'_{n}\in\{0,1\}$ implies either the absence ($0$) or presence ($1$) of the corresponding feature in $\mathbf{x}$, and $D$ is the feature dimension. The original model $f(\mathbf{x})$ is then approximated with an explanation model $g(\mathbf{x}')$:
        
        \begin{equation}
            \label{eq:SHAP}
            f(\mathbf{x})\approx g(\mathbf{x}') = \phi_{0}+ \sum_{n=1}^{D}\phi_{n}x'_{n} 
        \end{equation}
        where $\phi_{n}$ is the SHAP value for feature $x_{n}$, $\phi_{0}=f(h_x(\mathbf{0}))$ and $h_{x}$ is a mapping function which converts simplified features $\mathbf{x}'$ to original features $\mathbf{x}$, i.e., $\mathbf{x} = h(\mathbf{x}')$. The explanation model $g(\mathbf{x}')$ is trained to approximate the original model output $f(\mathbf{x})$ using the sum of SHAP values corresponding to the features for which $x'_n=1$ and coefficients $\phi_{n}$ in place of true SHAP values. SHAP values $\phi_n$ are derived for each class and can be positive and negative valued, indicating support (or not) for the corresponding class hypothesis. Further information and details are available in~\cite{SHAP, molnar_2020}.
    
    \begin{figure}[!t]
            \centering
            \begin{subfigure}{\columnwidth}
                \includegraphics[trim=38 0 0 0, scale=0.4]{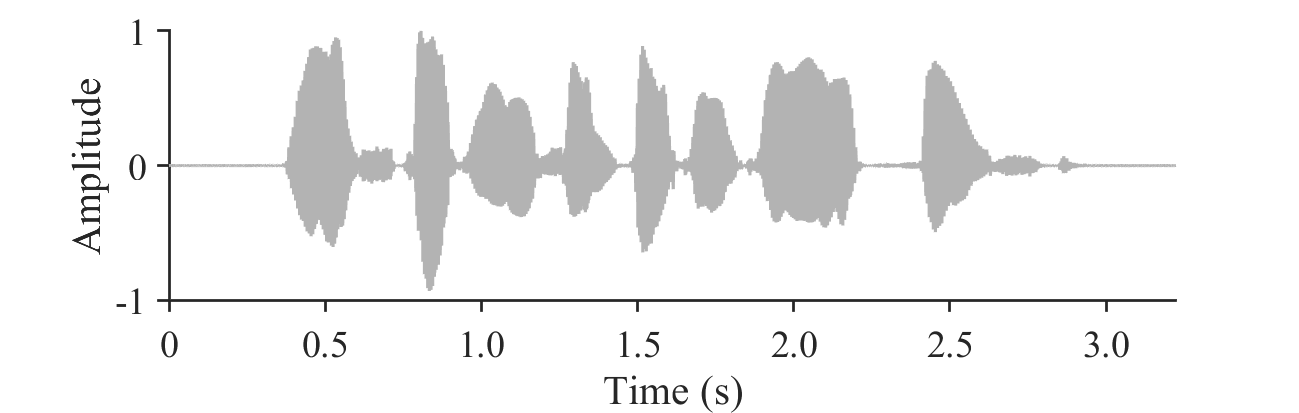}
                \caption{waveform}
                \label{fig:f1a}
                \end{subfigure}
            \begin{subfigure}{\columnwidth}
               \includegraphics[trim=38 0 0 0, scale=0.4]{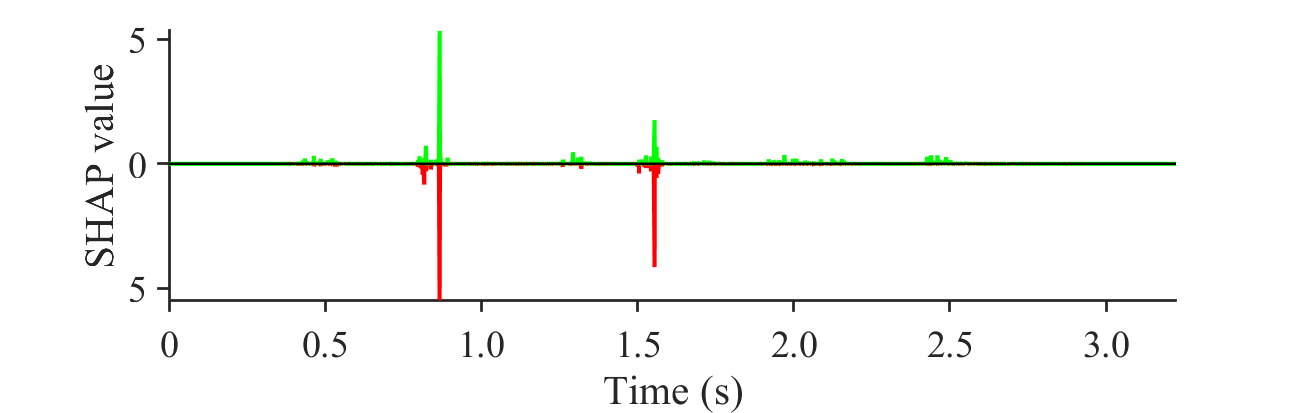}
                \centering
                \caption{SHAP values for bona fide (green) and spoofed (red)}
                \label{fig:f1b}
                \end{subfigure}
            
            \caption{Time waveform and SHAP values for the `LA\_T\_2909480’ A03 text-to-speech spoofed utterance `\emph{Well, Scotland had better grow up, fast}' from the ASVspoof 2019 LA database and a 1D-Res-TSSDNet spoofing detection model.}
            \label{fig:f1}
    \end{figure}

    \subsection{Example visualisations}
        Our first use of SHAP analysis applied to explainable spoofing detection~\cite{ge2021SHAP} was performed using spectro-temporal decompositions in the form of magnitude spectrograms.  This choice was motivated by their use to study easily the influence of different time intervals and frequency sub-bands upon the outputs of different classifiers, with the emphasis being upon characterising \emph{classifier behaviour}.
        
        The goal of the current work relates more to \emph{attack analysis}.  Attack artefacts, though, are undeniably the result of classifier behaviour; and they are inseparable within the current framework.  Accordingly, in a modest attempt to decouple or marginalise classifier influences, the work reported in this paper was also performed with two different classifiers.
    
        Nowadays, a growing number of competitive systems operate directly upon raw waveforms~\cite{cldnn, ma21d_rw_resnet,Teng2021ComplementingHF}. Since these approaches have potential to discover artefacts that might be lost through hand-crafted pre-processing or feature extraction, there is hence an interest to apply SHAP analysis to classifiers operating upon raw waveforms, in addition to spectro-temporal decomposition. Our expectation is that the use of two such different classifiers may help to provide a more generic account of spoofing artefacts that is less specific to the behaviour or limitations of a single feature representation and classifier alone.
    
    \subsubsection{Raw waveforms}
    
        The waveform of a spoofed utterance $x[n]$ selected from the A03 spoofed utterances in the ASVspoof 2019 LA training partition is depicted in Fig.~\ref{fig:f1a}. Using an arbitrary classifier which operates on the raw waveform, SHAP values $\phi_{n}$ are obtained using Eq.~\ref{eq:SHAP} for each sample $x[n]$ and each class: bona fide and spoofed.
        
        We have found that SHAP values for each class are approximately negatively symmetric, i.e.\ positive SHAP values for the bona fide class are correlated with negative values for the spoof class, and vice versa. To avoid redundancy, everywhere in this paper, we consider only positive SHAP values $\phi_n$ that favour either class. They are plotted in Fig.~\ref{fig:f1b} for the same utterance. Those for the bona fide class are illustrated in green whereas those for the spoof class are inverted and plotted below the abscissa in red.

        The plot shows that the classifier uses information distributed across the full utterance, though SHAP values are mostly larger for speech segments and lower for non-speech segments. The largest SHAP values are observed for speech segments around 0.9 seconds and 1.5 seconds which are most supportive of spoofed class hypotheses. Hence, while there is information everywhere in the utterance, the classifier output is dominated by cues or artefacts located in a specific temporal intervals.  

    \subsubsection{Magnitude spectrograms}
    
\begin{figure}[!t]
            \centering
            \begin{subfigure}{\columnwidth}
              \includegraphics[trim=38 0 38 0, scale=0.4]{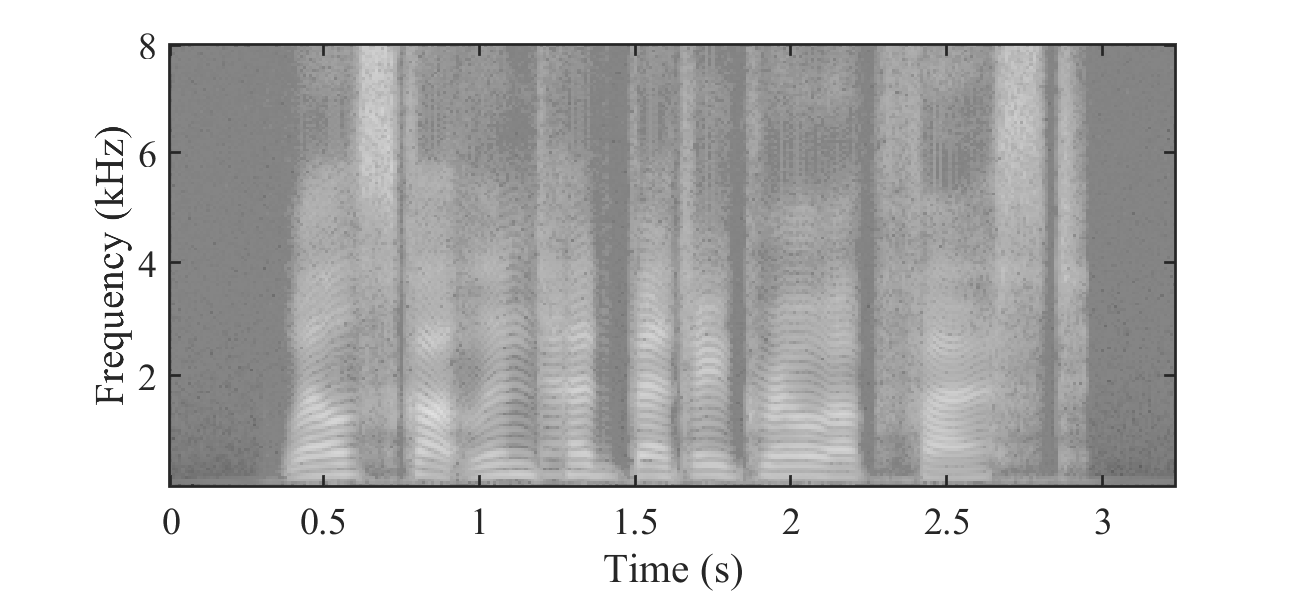}
                \centering
                \caption{magnitude spectrogram}
                \label{fig:f2a}
                \end{subfigure}
            \begin{subfigure}{\columnwidth}
            \includegraphics[trim=26 0 38 0, scale=0.4]{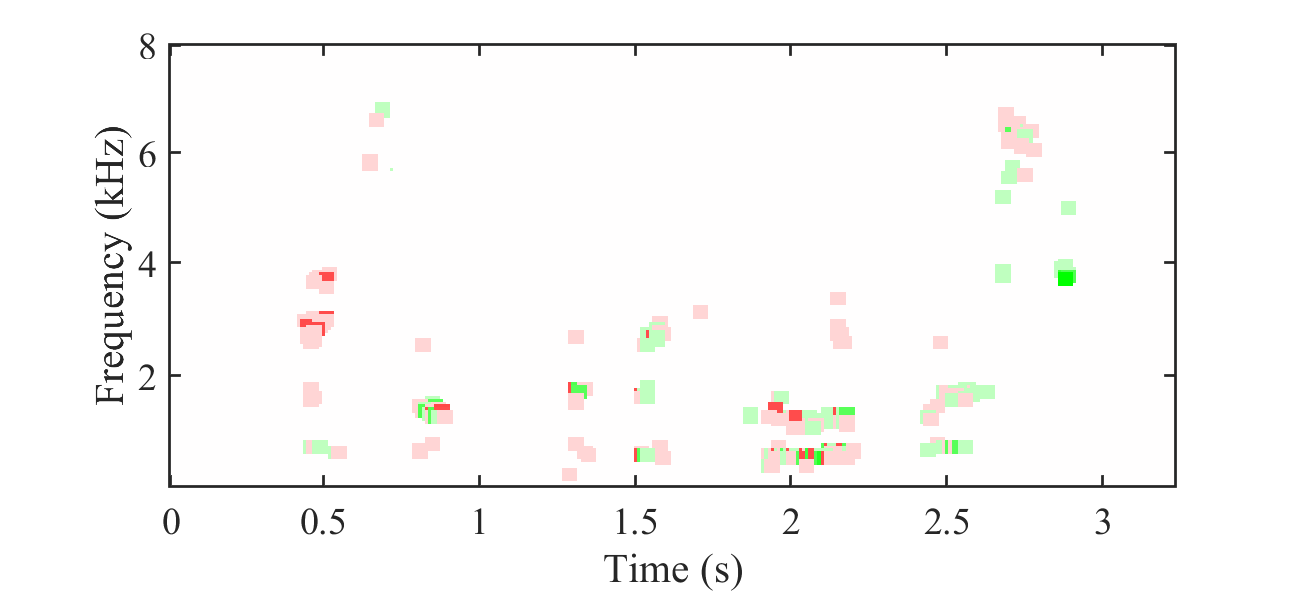}
                \centering
                \caption{SHAP values}
                 \label{fig:f2b}
                \end{subfigure}
            
            \caption{As for Fig.~\ref{fig:f1} except for a magnitude spectrogram input representation and a 2D-Res-TSSDNet spoofing detection model. SHAP values in (b) are dilated to improve their visualisation.}
            \label{fig:f2}
    \end{figure}

        The spectrogram of the same utterance as in Fig.~\ref{fig:f1a} is illustrated in Fig.~\ref{fig:f2a}.  In similar fashion to~\cite{sivasankaran21_SHAP_SE, ge2021SHAP,becker2018interpreting}, SHAP values $\phi_{(m,n)}$ for an arbitrary classifier which operates upon such spectro-temporal magnitude estimates $\mathbf{X}(m, n)$, where $m$ is the spectral bin and $n$ is the frame index, are illustrated in Fig.~\ref{fig:f2b}. Once again, we illustrate positive SHAP values only, which are now encoded in the colour intensity, still green for the bona fide class and red for the spoof class.
    
        As for the temporal plots in Fig.~\ref{fig:f1}, we find that information is scattered across the full spectrogram, though SHAP values are again greater for speech intervals than for non-speech intervals. Since the SHAP values are now encoded in the colour intensity, it is more difficult to identify from the visualisation the specific temporal or spectral intervals that have the greatest influence upon the classifier output, even if the most influential temporal segments appear to be different to those highlighted in Fig.~\ref{fig:f1}.
    
\section{Experimental setup}

    In this section we describe the classifiers used for the remainder of this work presented in this paper, specific implementation details and the pruning of SHAP values to better highlight the temporal and/or spectral intervals that bear the greatest influence upon the classifier output.

    \subsection{Spoofing detection classifiers}

        We used a pair of different classifiers, namely the 1D- and 2D-Res-TSSDNet systems proposed in~\cite{hua2021TSSDNet}.  
        The motivation for using these two particular systems are: 
        they are freely available as open source\footnote{\href{https://github.com/ghuawhu/end-to-end-synthetic-speech-detection}{https://github.com/ghuawhu/end-to-end-synthetic-speech-detection}} and can hence be used to reproduce our results;
        they share a similar network architecture consisting of several convolutional blocks with residual connections, followed by a global max pooling layer and three fully-connected layers, and both produce scores for bona fide and spoofed classes;
        they are efficient and can be trained quickly;
        their relatively uncomplicated structures should offer better potential for explainability.
        
        The 1D-Res-TSSDNet has only 0.35M trainable parameters whereas the 2D-Res-TSSDNet still has only 0.97M; our own solutions have a considerably greater number~\cite{ge21_raw_pc_darts, tak2021rawnet2}. The core difference between the two classifiers is that the 1D-Res-TSSDNet is based upon 1D convolutional operations applied initially to raw audio waveforms, whereas the 2D-Res-TSSDNet is based upon 2D convolutional operations applied to magnitude spectrograms 
        extracted using 20~ms Hamming windows with a 10~ms shift and a 320-point FFT.  
       
    \subsection{Implementation details}
        All experiments reported in this paper were conducted using the ASVspoof 2019 Logical Access (LA) database~\cite{asvspoof2019}.  The training partition is used to update network parameters whereas the evaluation partition is used as validation to select the best model. System performance is estimated using the equal error rate (EER). Since the goal of this work is to study spoofing artefacts and not to learn the best performing classifier, we make no use of the development partition for any experiments and use EER estimates only for the purposes of model selection. Both systems are optimised by minimising the weighted cross-entropy loss between spoofed and bona fide classes. Models are trained using the Adam optimiser~\cite{adam} with a learning rate of 0.001 and an exponential learning rate decay of 0.95. 
        
        Instead of using fixed-length inputs resulting from concatenation or truncation in order to create fixed-size mini-batches, both models operate upon variable, whole-length utterances. This choice was found to be beneficial in terms of explainability since it allows us to study more easily and consistently the influence upon classifier ouput of information available in specific temporal or spectral intervals. We found that concatenation in particular can result in a classifier making different use of the same information in repeated short intervals or sub-bands~\cite{ge2021SHAP},
        a behaviour which can complicate studies in explainability. 
       
        Variable-length input data is treated as described in~\cite{wang2021comparative} whereby training utterances of similar length are zero-padded to give uniformly-sized mini-batches.
        For experiments involving SHAP analysis, as opposed to model updating, utterances are nonetheless fed to the classifiers without any such padding. Though the use of variable-length inputs does degrade performance, we stress again that this work targets explainability, not performance. For reference, the EERs of the 1D-Res-TSSDNet and 2D-Res-TSSDNet are 6.90\% and 4.28\% respectively.

    \subsection{SHAP analysis and pruning}
                
        SHAP values\footnote{\href{https://captum.ai/api/gradient_shap.html}{https://captum.ai/api/gradient\_shap.html}} for a given input feature are computed from background features (also called baselines) of identically-sized zero-valued vectors. The number of randomly generated examples per sample is set to 20. All experiments reported in the paper are performed with the same random seed on a single NVIDIA GeForce RTX 3090 GPU. Results are reproducible with the same random seed and GPU environment using the same SHAP implementation and our scripts available online.\footnote{\href{https://github.com/GeWanying/shap-anti-spoofing}{https://github.com/GeWanying/shap-anti-spoofing}}

\begin{figure}[!t]
        \centering
        \begin{subfigure}[b]{0.525\columnwidth}
                \centering
                \includegraphics[width=\columnwidth]{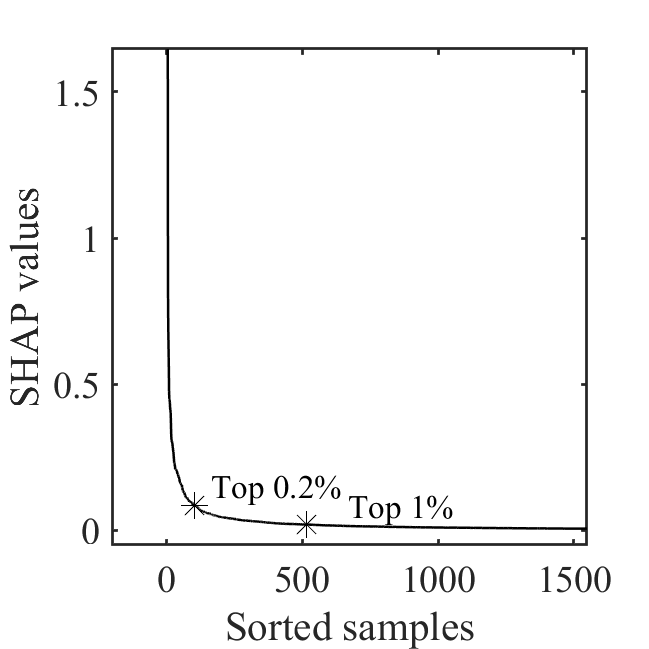}
                \caption{waveform feature}
                \label{fig:sorted_1D}
        \end{subfigure}
        \hfill
        \begin{subfigure}[b]{0.525\columnwidth}
                \centering
                \includegraphics[width=\columnwidth]{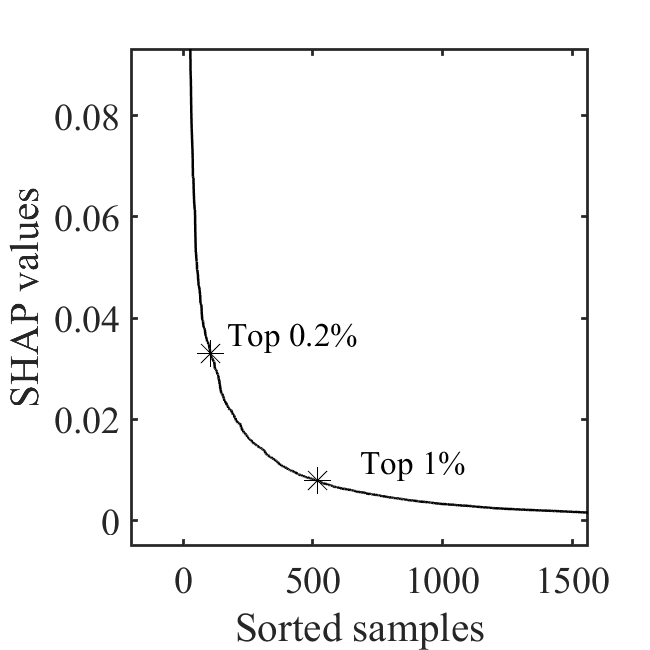}
                \caption{spectrogram feature}
                \label{fig:sorted_2D}
        \end{subfigure}
        \caption{Histograms of SHAP values for both classifiers and the utterance shown in Fig.~\ref{fig:f1}.}
        \label{fig:Sorted}
        \vspace{-0.32cm}
    \end{figure}

        While SHAP values are derived for every time sample $n$ or every specto-temporal sample $[m,n]$, we observed particularly high SHAP values for typically small numbers of specific samples or intervals. Histograms of SHAP values for both 1D and 2D classifiers depicted in Fig.~\ref{fig:Sorted} show approximate negative exponential distributions and that the highest SHAP values are accounted for by approximately 0.2\% of samples.  By selectively pruning the input, we confirmed through experiments not reported here that samples with these highest SHAP values have a dominant influence upon the classifier output whereas samples with near-to-zero SHAP values have comparatively less influence. Accordingly, all SHAP visualisations reported in the remainder of this paper correspond to the 0.2\% of samples with the highest, dominant SHAP values. In similar fashion to~\cite{becker2018interpreting}, SHAP values in support of the spoofed class (red only) are encoded in colour intensity superimposed upon waveform or magnitude spectrogram plots.

\section{Attack analysis}

    We present an attack analysis which focuses upon spoofing attacks contained in the ASVspoof 2019 LA training partition.
    It includes spoofed utterances generated with 6 different algorithms including 4 TTS attacks (A01-A04) and 2 VC attacks (A05-A06).
    \begin{table*}[!t]
        \caption{\label{table:train_attacks} {\it Artefact description of attacks in ASVspoof 2019 LA train partition.}}
            \centering \renewcommand{\arraystretch}{1.25}
            {%
            \begin{tabular}{|p{1cm}|p{1.5cm}|p{5cm}|p{5cm}|}
            \hline
            \multicolumn{2}{|c|}{} & \multicolumn{2}{c|}{Found artefacts}\\\hline
            Attack & Algorithm & Waveform & Spectrogram\\
            \hline
            A01 & TTS & Vowels & Lower frequency bands, leading 0.5s \\  \hline
            A02 & TTS & Single dominant vowel & Lower \& higher frequency bands, unvoiced \textbackslash s\textbackslash  \\\hline
            A03 & TTS & Less densely distributed in vowels  & Lower frequency bands  \\ \hline
            A04 & TTS & Non-speech, low energy speech segments (voice onsets and offsets) & Full spectrum, unvoiced speech, clicks \\ \hline 
            A05 & VC & Voice onset, vowels & Full spectrum, higher energy formant frequencies\\\hline
            A06 & VC & Speech distortion & Lower frequency bands  \\ \hline 
            \end{tabular}
                    }
            \end{table*}
    \begin{figure*}[!hp]
            \centering
            \begin{subfigure}{\columnwidth}
                \includegraphics[trim=38 0 0 0, scale=0.4]{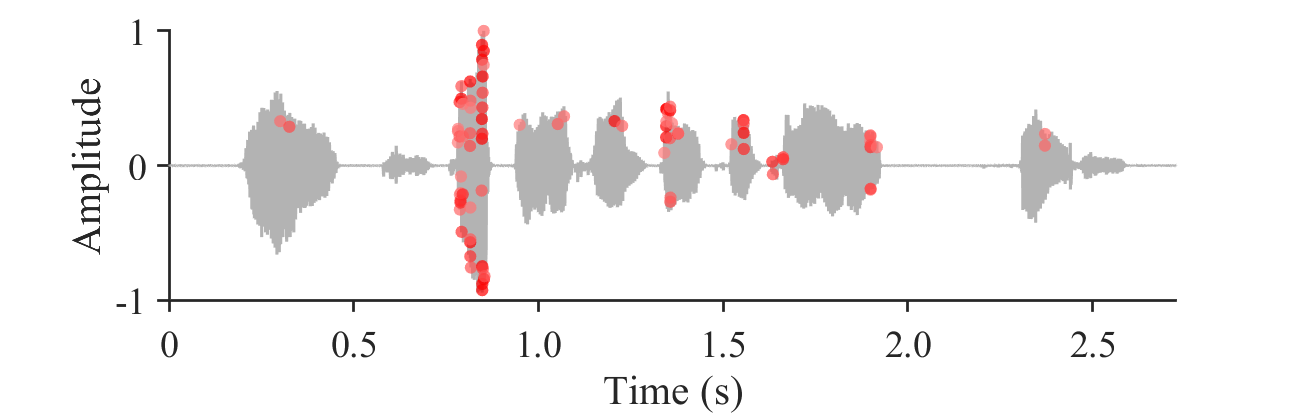}
                \includegraphics[trim=88 0 0 0, scale=0.19]{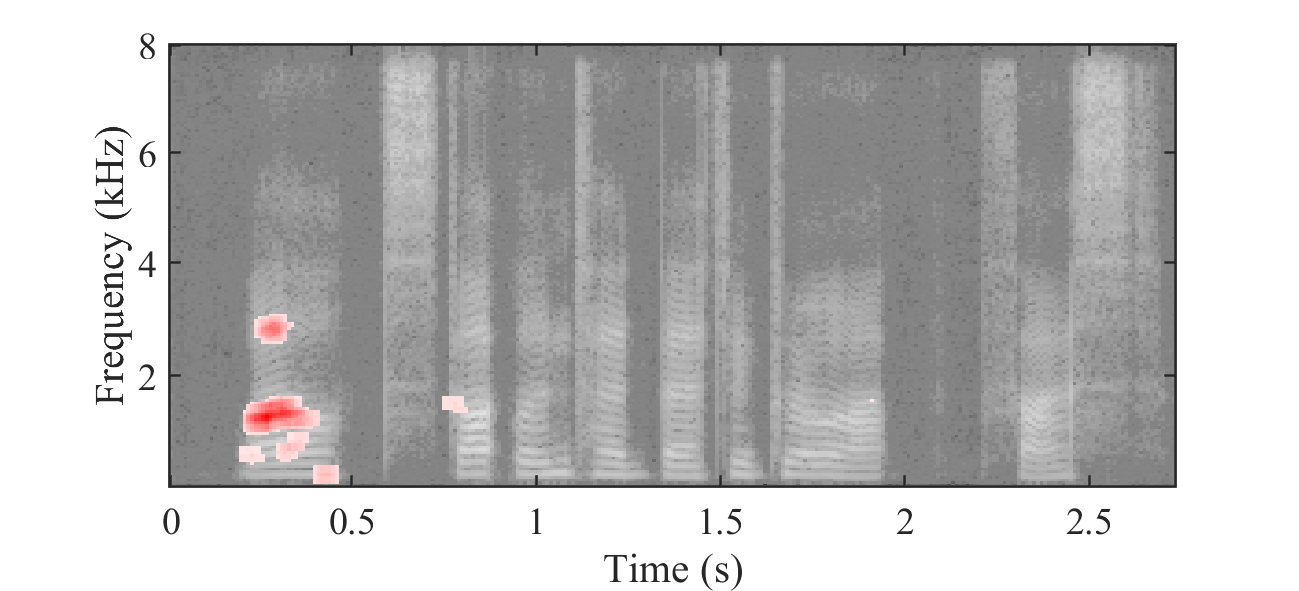}
                \caption{A01 attack utterance LA\_T\_3566209}
                \label{fig:A01}
                \end{subfigure}
            \begin{subfigure}{\columnwidth}
                \includegraphics[trim=38 0 0 0, scale=0.4]{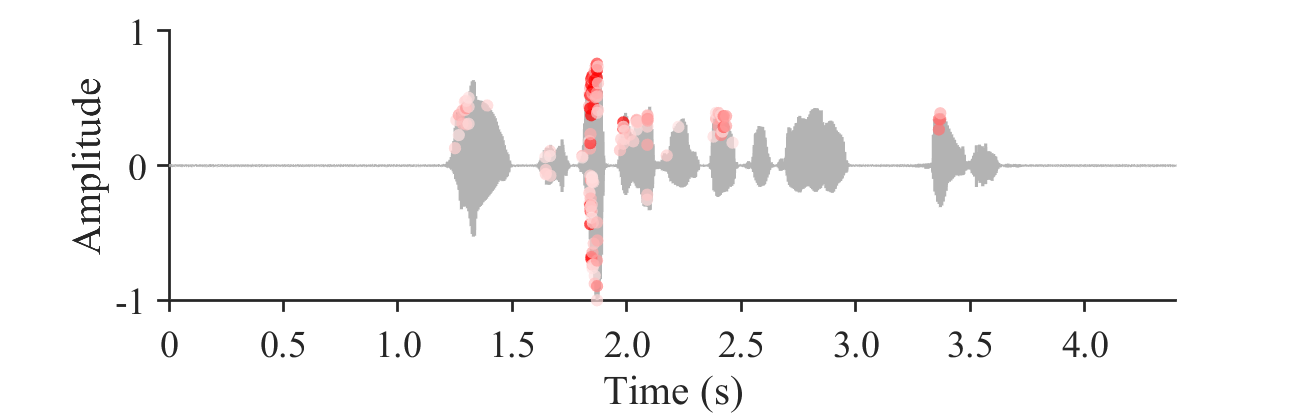}
                \includegraphics[trim=88 0 0 0, scale=0.19]{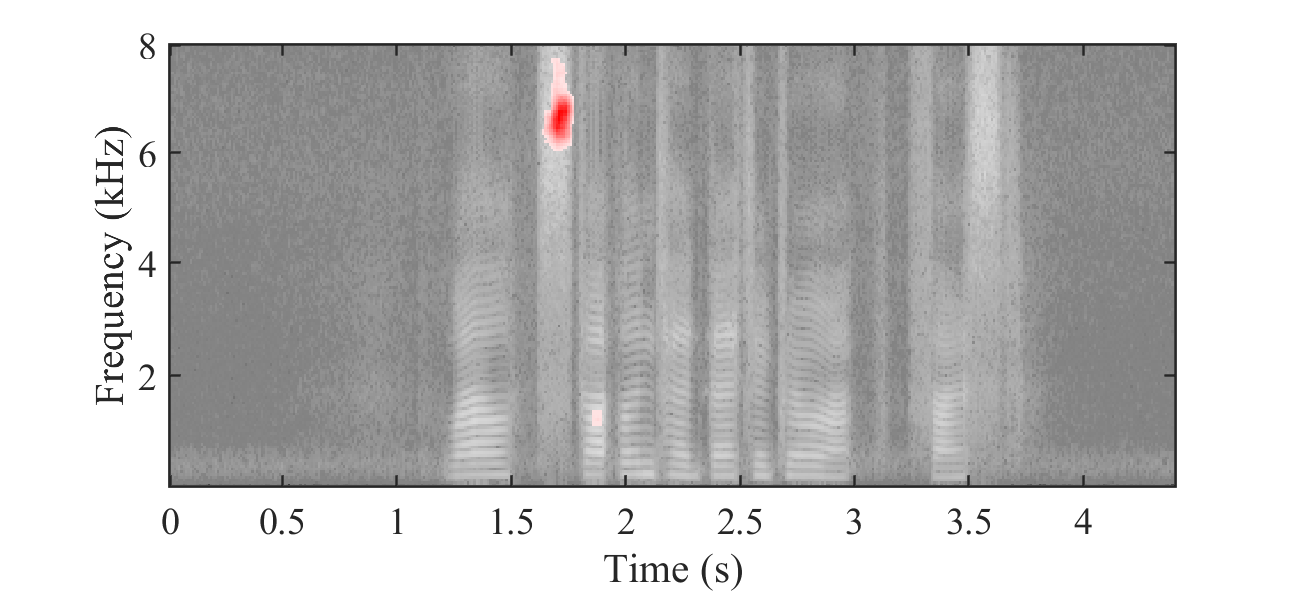}
                \caption{A02 attack utterance LA\_T\_1590397}
                \label{fig:A02}
                \end{subfigure}

            \begin{subfigure}{\columnwidth}
                \includegraphics[trim=38 0 0 0, scale=0.4]{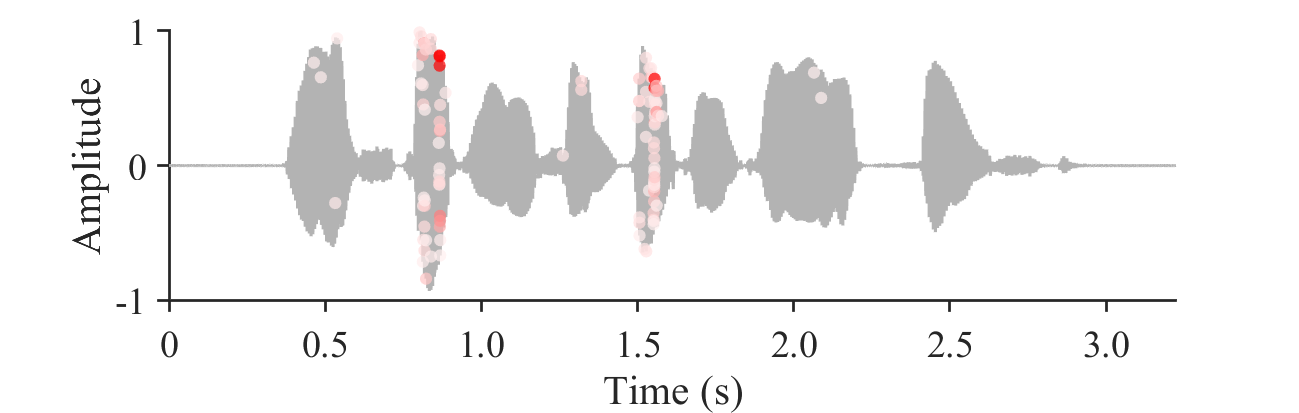}
                 \includegraphics[trim=88 0 0 0, scale=0.19]{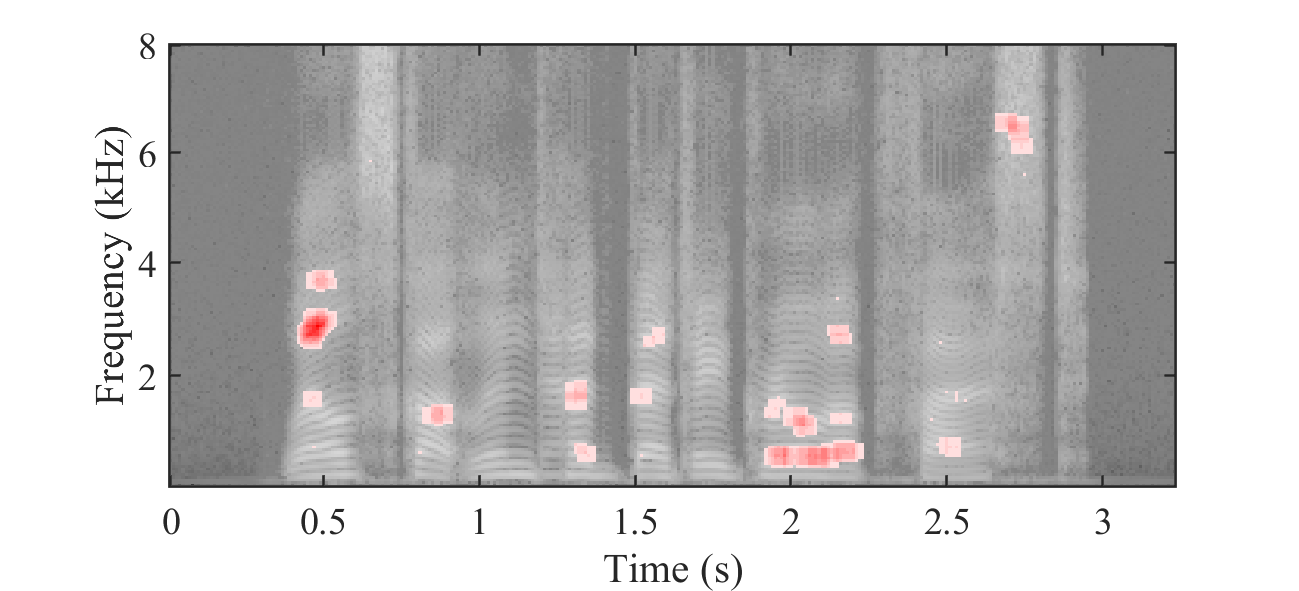}
                \caption{A03 attack utterance LA\_T\_2909480}
                \label{fig:A03}
                \end{subfigure}
            \begin{subfigure}{\columnwidth}
                \includegraphics[trim=38 0 0 0, scale=0.4]{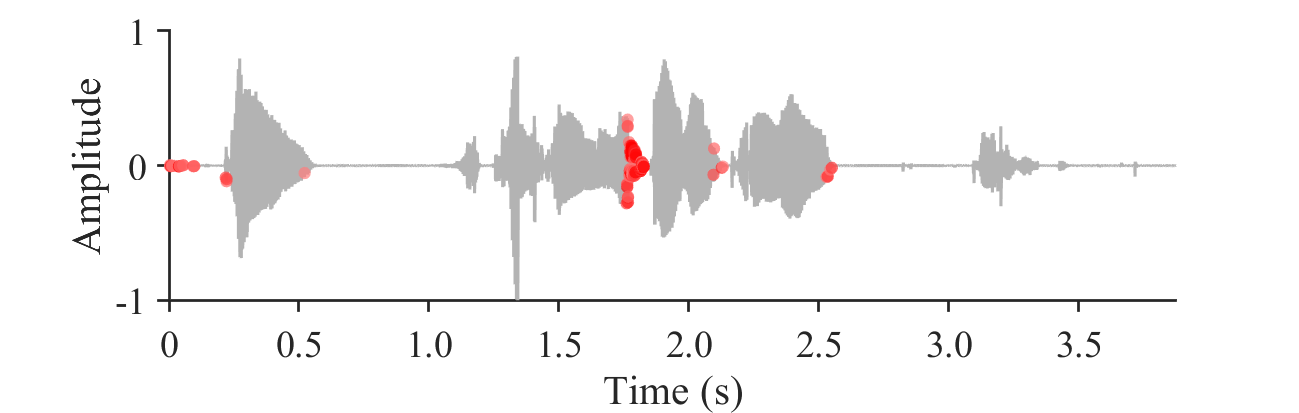}
                \includegraphics[trim=88 0 0 0, scale=0.19]{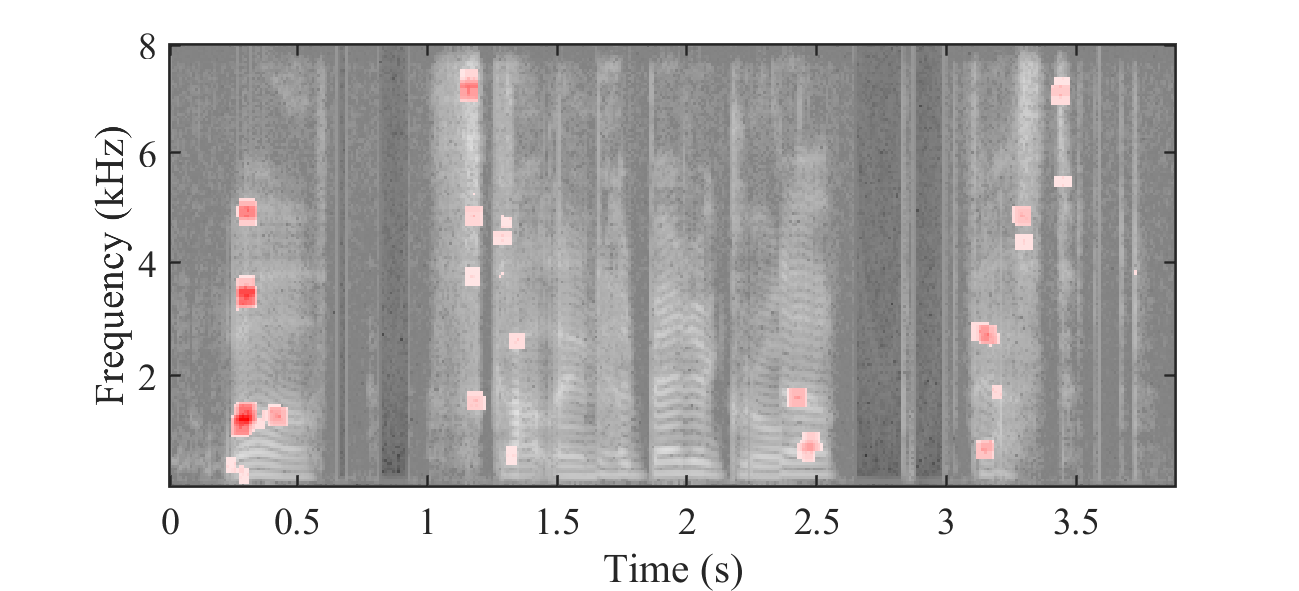}
                \caption{A04 attack utterance LA\_T\_5116902}
                \label{fig:A04}
                \end{subfigure}

            \begin{subfigure}{\columnwidth}
                \includegraphics[trim=38 0 0 0, scale=0.4]{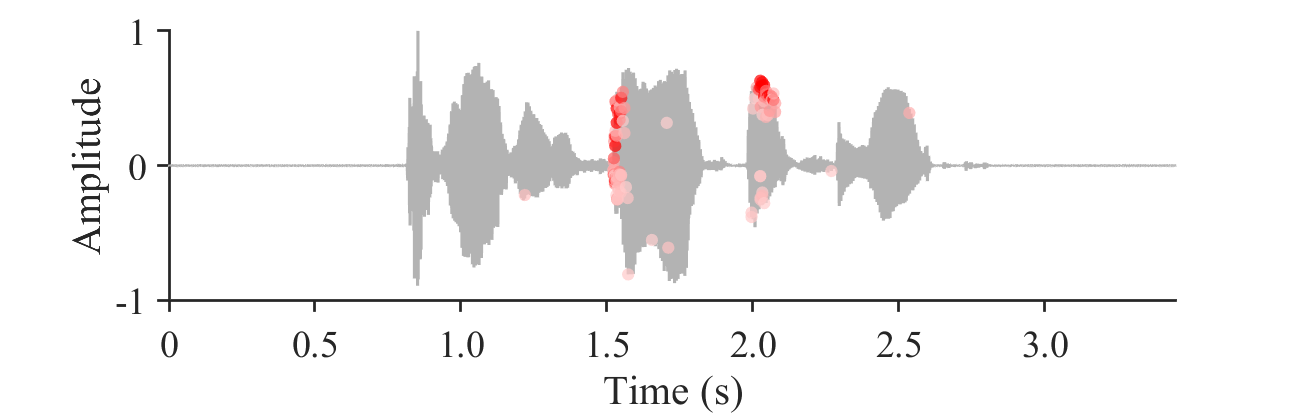}
                \includegraphics[trim=88 0 0 0, scale=0.19]{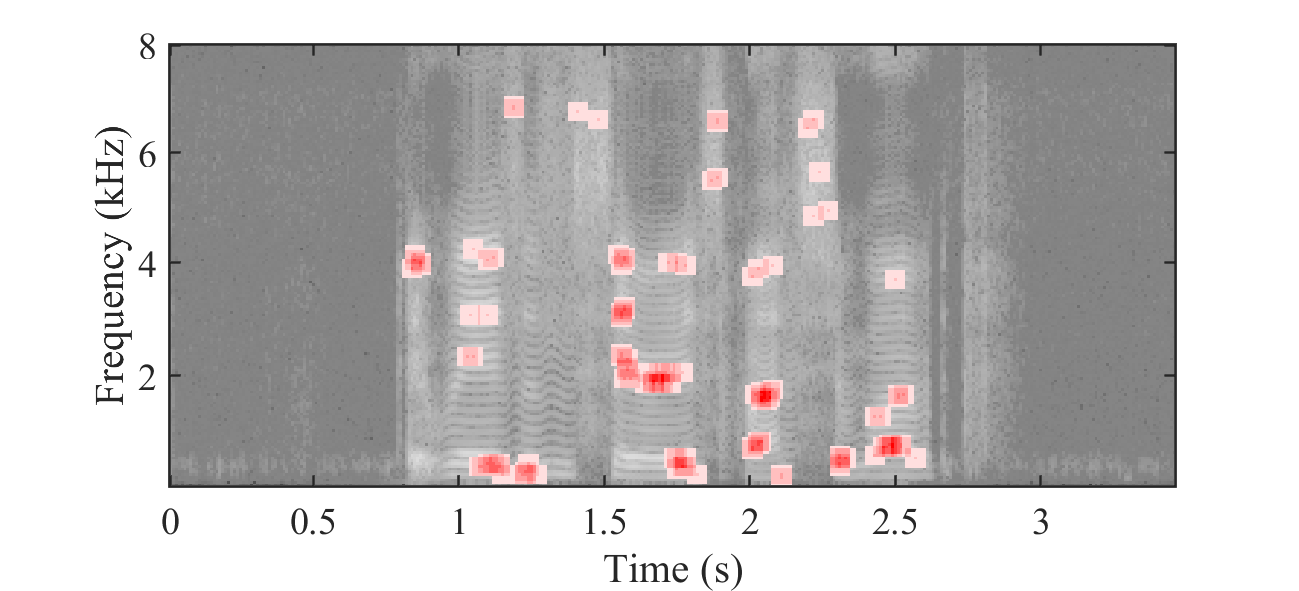}
                \caption{A05 attack utterance LA\_T\_3134909}
                \label{fig:A05}
                \end{subfigure}
            \begin{subfigure}{\columnwidth}
                \includegraphics[trim=38 0 0 0, scale=0.4]{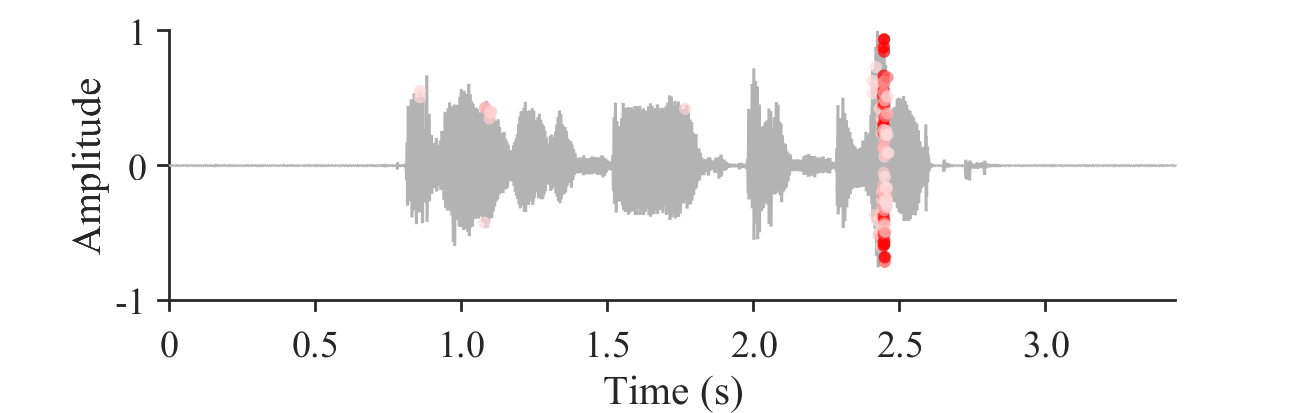}
                \includegraphics[trim=88 0 0 0, scale=0.19]{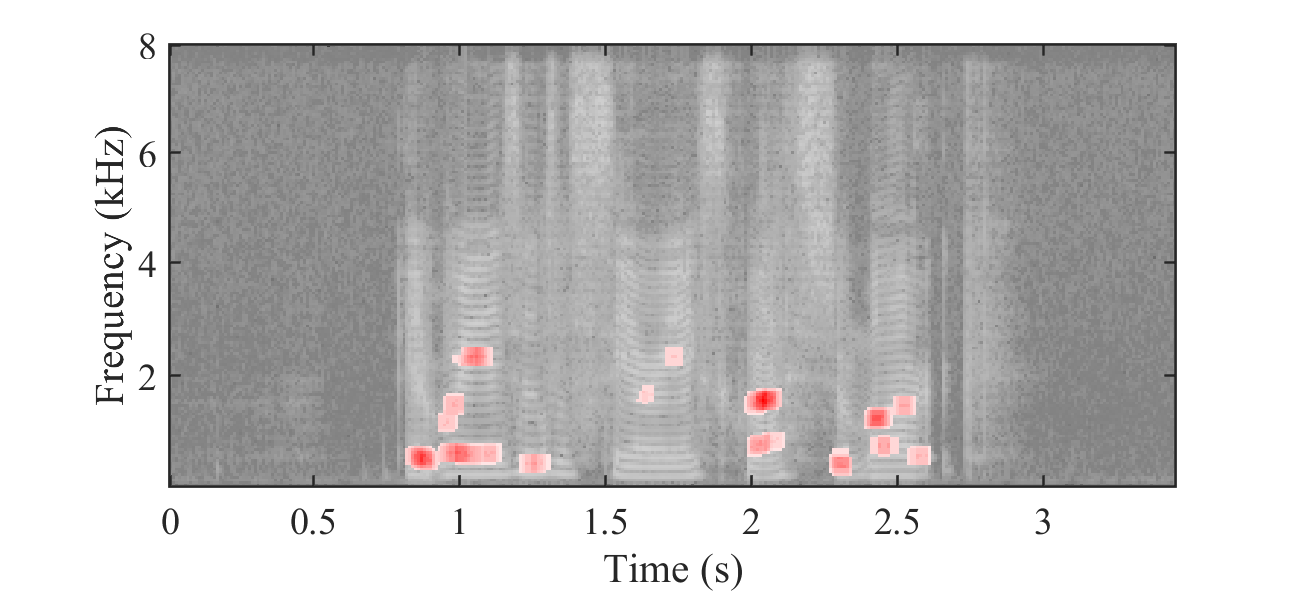}
                \caption{A06 attack utterance LA\_T\_5300749}
                \label{fig:A06}
                \end{subfigure}

            \caption{SHAP values for the A01-A04 utterance `\emph{Well, Scotland had better grow up, fast}' and A05 and A06 utterance `\emph{It raises a serious question mark}'. SHAP values superimposed upon either waveforms or spectrograms and dilated in case of the latter, in similar fashion to Fig.~\ref{fig:f2}.}
            \label{fig:train_attacks}
            \end{figure*} 
        For all reported results, the classifiers are trained using bona fide data and \emph{matched} spoofed data only -- SHAP analysis for an A01 attack utterance is performed using classifiers trained only with A01 attack utterances, not the full set of A01-A06 training attack utterances. We found that this approach helps in the identification of consistent artefacts for each attack. Our aim is to isolate or marginalise the influence of classifier behaviour to the extent that this is possible. We accept that, within our current framework, it is impossible to \emph{fully} isolate classifier behaviour from attack artefacts, since the latter are always learned using a specific classifier. By training the classifier with matched attack utterances, we hope at least to observe more clearly and consistently the artefacts that are specific to the given attack, and not those learned from other attacks. This approach is consistent with our goal of \emph{attack analysis}.
    
        The following analyses were preformed upon a random selection of 100 utterances for each training attack. Example results are illustrated in Fig.~\ref{fig:train_attacks} which shows both waveforms and spectrograms with the corresponding superimposed 0.2\% highest intensity-encoded SHAP values. The illustrated examples are specifically chosen to use the same utterance for TTS attacks and the same utterance for VC attacks\footnote{Text inputs used in generating TTS attack utterances are not necessarily consistent with the text content of VC attacks~\cite{wang2020asvspoof2019}.}. A discussion of results for each attack is presented in the following whereas a summary of the principal, consistent artefacts observed in each case is presented in Table~\ref{table:train_attacks}.  These were identified from SHAP analysis and casual listening test.

        \textbf{A01} is a neural network (NN) based TTS attack with a WaveNet~\cite{oord2016_wavenet} vocoder.  A waveform and spectrogram with superimposed SHAP values are illustrated in Fig.~\ref{fig:A01}. We observed differences for 1D and 2D classifiers. For the 1D classifier we found most artefacts in vowel segments though we could not identify a particular vowel for which SHAP values are consistently the highest. Most artefacts are found at low frequency bands. For a substantial number of utterances, the 2D classifier identifies artefacts in the leading 0.5 seconds of speech. This might be the result of A01 attacks having a consistently shorter leading silence interval compared to  bona fide utterances~\cite{muller21_silence}.

        \vspace{0.2cm}
        \textbf{A02} is also an NN-based TTS attack, but with a WORLD~\cite{morise2016_WORLD} vocoder. Results are illustrated in Fig.~\ref{fig:A02}. Like the A01 attack, the 1D classifier finds artefacts in vowel segments, such as the \textbackslash o\textbackslash\ vowel in the given example. The 2D classifier finds artefacts mostly at lower frequencies and also at higher frequency bands above 6~kHz. For A02 attacks and the 2D classifier, consistent artefacts are identified for the unvoiced sound \textbackslash s\textbackslash segments. 

        \vspace{0.2cm}
        \textbf{A03} is a different NN-based TTS attack also with a WORLD vocder. Results are shown in Fig.~\ref{fig:A03}. For the 1D classifier, artefacts are found mostly in vowel segments, but are less densely distributed compared to A01 and A02 (relatively fewer dark-red points). The reason might be that artefacts in A03 attacks are located in particular samples which are different with the neighborhood ones. While we found that artefacts lie mostly at lower frequencies in the case of the 2D classifier, we did not succeed in identifying artefacts within consistent speech segments.
    
        \vspace{0.2cm}
        \textbf{A04} is a waveform concatenation TTS attack. Results are shown in Fig.~\ref{fig:A04}.
    For some A04 attacks we found artefacts to lie within leading non-speech and low energy speech segments (onsets and offsets of speech), a characteristic which differentiates A04 from the other TTS attacks.  These observations may correspond to the use of waveform concatenation and may explain why such attacks generated with the MaryTTS platform~\cite{schroder2011_marytts} were initially challenging to detect~\cite{asvspoof2015}. The 2D classifier uses cues throughout the full spectrogram. Other consistent artefacts were found in unvoiced segments and click sounds (around 1.2 and 3.5 seconds for the given example).
    
        \vspace{0.2cm}
        \textbf{A05} is an NN-based VC attack. Results are shown in Fig.~\ref{fig:A05}. In addition to more dominant artefacts in vowels segments (around 2.0 seconds in the given example), the 1D classifier also finds consistent artefacts in lower energy voice onset segments (around 1.6 seconds). Similar to the TTS A04 attack, the 2D classifier finds artefacts across the full spectrum, rather than specific sub-bands. Nonetheless, higher SHAP values correspond generally to lower frequencies and higher energy formant frequencies (around 2 and 4 kHz).
    
        \vspace{0.2cm}
        \textbf{A06} is transfer-function-based VC attack~\cite{matrouf2006}. Results are shown in Fig.~\ref{fig:A06}. For the 1D classifier, we observed temporal intervals with high SHAP values to correspond to noticeably distorted, speech sounds.  These seem to correspond to variations in unnaturally high velocity. Using the 2D classifier, artefacts are found mostly at lower frequencies below 3 kHz. There were not found to correspond consistently to any particular speech sounds other than high-energy segments and neither do they correspond consistently to the distortions identified using the 1D classifier.

\section{Discussion and conclusion}

    Reported in the paper is our use of SHAP analysis to characterise the six synthetic speech and converted voice spoofing attacks that make up the training partition of the ASVspoof 2019 LA database. The analysis, performed with a pair of classifiers operating upon either time-domain waveforms or spectro-temporal magnitude estimates show the importance of analysing artefacts using different representations and classifiers; they identify different artefacts. While the artefacts are also attack-specific,  we nonetheless observed consistencies in attack classes and algorithms.
    
    Three of the four synthetic speech attacks exhibit temporally-distinct artefacts located within vowel segments.  These three attacks are all neural network based text-to-speech algorithms.  The fourth synthetic speech attack, by contrast a waveform concatenation based approach, exhibits different artefacts located within lower energy segments.  Differences in artefacts are possibly indicative of the spoofing attack algorithm. Differences were also observed for the two voice conversion algorithms, with one showing artefacts during voice onset segments and distributed throughout the spectrum, and the other corresponding to notable distortions and lower frequencies. There are, however, some consistencies between the artefacts produced by both converted voice and synthetic speech attacks, particularly artefacts in low frequency bands.
    While differences in artefacts may paint an ominous picture for the designers of spoofing detection systems, consistencies are encouraging since they offer some insights as to why, as shown in broader literature, classifiers trained using spoofed utterances generated using only a small set of spoofing algorithms are nonetheless effective in the face of utterances generated with different spoofing algorithms.  
    
    Limitations of the current work relate to the use of classifiers trained using matching spoofed utterances.  This luxury is unavailable for practical use cases for which the nature of spoofing attacks is unknown and unpredictable. Explainability for \emph{ideal} scenarios involving \emph{known} attacks is already extremely challenging. Even in these conditions the identification of consistent trends is difficult. We are now working to apply the same technique to different, \emph{unknown} spoofing attacks in a manner that nonetheless identifies consistent artefacts. This work, using utterances contained in the ASVspoof 2019 LA evaluation partition generated by 13 different attacks, is proving even more challenging. 
    
    One current direction for our future work involves the identification of attack-specific artefacts using matched training data and then their re-identification using the same attack-specific classifiers used in this work. Explainable spoofing detection solutions could then make use of SHAP visualisations as presented in this paper, but also comparisons to artefacts previously identified for known attack algorithms. Another direction for future work involves the linking of artefacts to specific attack algorithm components, e.g.\ a particular representations, neural architectures or vocoders.
    
\section{Acknowledgements}
    
    This work is supported by the TReSPAsS-ETN project funded from the European Union’s Horizon 2020 research and innovation programme under the Marie Skłodowska-Curie grant agreement No.860813. It is also supported by the ExTENSoR project funded by the French Agence Nationale de la Recherche (ANR).

\balance 
\bibliographystyle{IEEEbib}
\bibliography{references.bib}

\end{document}